\documentclass[conference]{IEEEtran}
\usepackage{booktabs} 
\usepackage{graphicx}
\usepackage{float}
\usepackage{dblfloatfix}
\usepackage{subcaption}
\usepackage{bm}
\usepackage{threeparttable} 
\usepackage[bookmarks=false]{hyperref}
\usepackage{subfloat}
\usepackage{comment}
\usepackage{etoolbox}
\usepackage{multirow}
\usepackage{array}
\usepackage{makecell}
\usepackage[sort,square,compress,comma,numbers]{natbib}
\usepackage{blindtext}
\usepackage[hang]{footmisc}
\usepackage[export]{adjustbox}
\usepackage{mathtools}
\usepackage{amsmath}
\usepackage{chngcntr,etoolbox} 
\usepackage[thinlines]{easytable}
\usepackage{enumitem}
\usepackage{xcolor} 
\usepackage{balance} 
\usepackage{tikz}
\usepackage{soul}

\cspreto{section}{\addtocounter{section}{-1}\refstepcounter{section}}
\cspreto{subsection}{\addtocounter{subsection}{-1}\refstepcounter{subsection}}

\newcommand{\RR}{\textsc{RetroRank}\xspace}  

\newcommand*\circled[1]{\tikz[baseline=(char.base)]{
            \node[shape=circle,draw,text=black, scale=0.65, fill=none,fill opacity=2, inner sep=0.3pt] (char) {#1};}}

\definecolor{Gray}{gray}{0.9}

\makeatletter
\renewcommand\footnoterule{%
  \kern-3\p@
  \hrule\@width.4\columnwidth
  \kern2.6\p@}
\makeatother
    
  
\begin{document}
\title{Recommending Bug-fixing Comments from Issue Tracking Discussions in Support of Bug Repair}

\author{
Rrezarta Krasniqi \\
Dept. of Computer Science and Engineering \\
University of Notre Dame \\
Notre Dame, IN, USA \\
rkrasniq@nd.edu
}

\maketitle

\begin{abstract}
In practice, developers search for related earlier bugs and their associated discussion threads when faced with a new bug to repair. Typically, these discussion threads consist of comments and even \emph{bug-fixing comments} intended to capture clues for facilitating the investigation and root cause of a new bug report. Over time, these discussions can become extensively lengthy and difficult to understand. Inevitably, these discussion threads lead to instances where bug-fixing comments intermingle with seemingly-unrelated comments. This task, however, poses further challenges when dealing with high volumes of bug reports. Large software systems are plagued by thousands of bug reports daily. Hence, it becomes time-consuming to investigate these bug reports efficiently. To address this gap, this paper builds a ranked-based automated tool that we refer it to as \RR. Specifically, \RR recommends bug-fixing comments from issue tracking discussion threads in the context of user \emph{query relevance}, the use of \emph{positive language}, and \emph{semantic relevance} among comments. By using a combination of Vector Space Model (VSM), Sentiment Analysis (SA), and the TextRank Model (TR) we show how that past fixed bugs and their associated bug-fixing comments with relatively positive sentiments can semantically connect to investigate the root cause of a new bug. We evaluated our approach via a synthetic study and a user study. Results indicate that \RR significantly improved performance when compared to the baseline VSM.
\end{abstract}

\section{Introduction}
\label{sec:introduction}
\emph{Bug-fixing comments} are comments posted in a discussion thread attached to a bug report. Typically, they provide bug-fixing clues such as: resolution instructions, inline links to related bug reports, commit links, patch links, and positive affirmation when a bug report has been fixed~\cite{noyori2019good,arya2019analysis,zhou2012should}. Examples of such clues are observed in the comments depicted in Figure~\ref{fig:bug_report_unresolved_b}. Some of these comments may include insightful details and display positive language used by the person providing a bug-fixing comment. In general, these bug-fixing comments can be found from previously fixed bugs.

Detecting bug-fixing comments from previously fixed bugs are important to program comprehension and maintenance because developers generally refer to similar solutions to understand the behavior of the current bug. As Bettenburg~\emph{et al.}~\cite{bettenburg2008duplicate} pointed out, descriptions of earlier solved bugs help developers to repair newly-reported bugs, and therefore can be beneficial for developers to understand the conditions under which a bug can be reproduced. The typical strategy to find bug-fixing comments is to either i) use a text web search engine, or ii) use a duplicate bug report detection technique. Duplicate bug report detection has the advantage of high customization to the problem domain, but also has the disadvantage of requiring domain-specific information such as complete bug reports or stack traces. Developers can turn to text-based search engines when they only have a short query in mind, but the results are likely to be lower quality due to the more general nature of the search problems that search engines are designed to solve. 

We have observed that bug-fixing comments tend to have two characteristics in common. First, bug-fixing comments include a description of the bug repair that often summarizes the bug behavior. Second, bug-fixing comments tend to be written in a more positive style than the remainder of the conversations. The pattern seems to be frustration for several comments, followed by elation when the bug is fixed. These emotions are reflected in the language used by the participants in the conversations. Several studies also reveal that sentiment patterns were noticeable in bug report discussion threads~\cite{maalej2015bug}, IT support tickets~\cite{blaz2016sentiment}, and GitHub issue comments~\cite{imtiaz2018sentiment}. 

Based on these observations, this paper builds an efficient ranked-based technique that we refer it as \RR. The core functionality of \RR is to recommend bug-fixing comments from discussion threads of past resolved bugs to address a new bug. These comments are ranked in the context of user \emph{query relevance}, use of \emph{positive language}, and \emph{semantic relevance}. Specifically, \RR combines the Sentiment Analysis (SA) and the TextRank Model (TR) with the Vector Space Model (VSM) as employed by Zhou et \emph{al}.~\cite{zhou2012should}. We use SA to detect positive sentiment in comments. We use the TR to capture the semantic connection amongst comments when lexical matching amongst related comments fails to be detected by the VSM model. Our three-pronged approach retrieves related bug-fixing comments, ultimately facilitating steps on addressing a new bug as denoted in Figure~\ref{fig:bug_report_unresolved_b}. 

We perform two distinct studies to evaluate our approach synthetic and user study. In the synthetic study, we comparatively evaluate different approaches, namely, VSM (baseline), VSM+SA (our approach with TR disabled), VSM+TR (our approach with SA disabled), and VSM+SA+TR used by \RR. Results indicate that VSM+SA and VSM+TR both outperformed the baseline VSM, but found that VSM+SA+TR was the best performing approach overall. In the user study, we evaluated best performing approach (i.e., VSM+SA+TR) to the baseline VSM. Our choice for using VSM as a baseline is supported by four premises: i) VSM can determine term relevance in bug report queries~\cite{ramos2003using},
ii) each discussion comment can be represented into a vector of term frequency-inverse document frequency (TF$\times$IDF) weights that can be enhanced for relevance ranking~\cite{paik2013novel}, iii) VSM can be enhanced by NLP methods when VSM alone fails to capture textual matching between queries and comments~\cite{WangZXAS08}, and iv) it is regarded as a suitable method for bug localization tasks~\cite{wang2014compositional,thung2014dupfinder,wang2011concern}. Results indicate that \RR when using combination VSM+SA+TR outperformed the baseline VSM. 
\section{Problem and Scope}
\label{sec:motivation}

One of the challenges directly affecting developer's productivity in the process of fixing bug reports is the time required to search and evaluate high volumes of reported bugs. Large software systems are plagued by thousands of bug reports daily~\cite{anvik2005coping}. As this number keeps increasing, the time needed to investigate these bug reports by developers becomes almost an unattainable task to achieve~\cite{marks2011studying}. Additionally, some of the bug reports are more complex than others. They contain discussion threads that are lengthy and frequently become difficult to understand as different stakeholders engage in extensive discussions~\cite{arya2019analysis}. Lengthy discussions pose substantial challenge to developers who must reason about which comments in a discussion thread are relevant to finding a potential solution for the bug report that they are investigating. 

To automate this process, our tool effectively and efficiently recommends the most relevant bug-fixing comments. To demonstrate the output of our tool, we provide an example. In Figure~\ref{fig:bug_report_unresolved_a}, we show an unresolved LibreOffice bug with ID \#34600. In Figure~\ref{fig:bug_report_unresolved_b}, we show a collection of related bug-fixing comments recommended from lengthy discussion threads from four various resolved past bugs. These recommended comments fall under the common theme of bug \#34600. For example, `comment $3$' extracted from discussions of bug \#34436 provides concrete suggestion on what condition the problem could be eliminated. Similarly, `comment $2$' extracted from discussions of bug \#33662 provides directions on how to handle the $90$ degree alignment issue. While `comment $4$' extracted from discussions of bug \#34136 and `comment $2$' of bug \#32795, confirm the fix of bug \#33622. 

Note that by using plain keyword-matching heuristics for extracting comments with keywords `fix' or `fixed' would be an insufficient search  criterion, as there exist other insightful comments that could capture sentiment features and rich patterns that are more affirmative~\cite{bahrainian2015sentiment}. Such insightful information is present in the comments from bugs \#33622 and \#34436, both characterized by positive sentiment and semantic relevance. This sentiment pattern was detected by using Sentiment Analysis. The semantic connection between descriptions of comments was obtained by using TextRank. 
\section{Background}
\label{sec:background}

This section describes three supporting technologies for our work, namely, VSM Model~\cite{salton1975vector}, TextRank Model (TR)~\cite{mihalcea2004textrank}, and the use of Sentiment Analysis (SA)~\cite{pang2008opinion}. 

\begin{subfigures}
\begin{figure}[!t]
\begin{center}
\includegraphics[width=6.8cm]{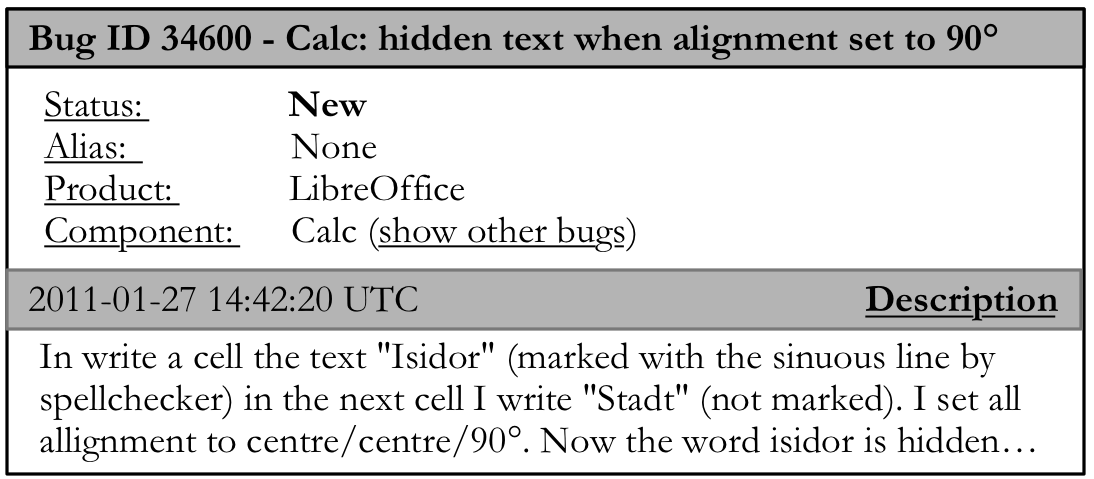}
\vspace{-0.20cm}
\caption{\label{first} Example of unresolved LibreOffice bug \#34600.}
\label{fig:bug_report_unresolved_a}
\vspace{-0.40cm}
\end{center}
\end{figure}

\begin{figure}[!t]
\begin{center}
\includegraphics[width=6.8cm]{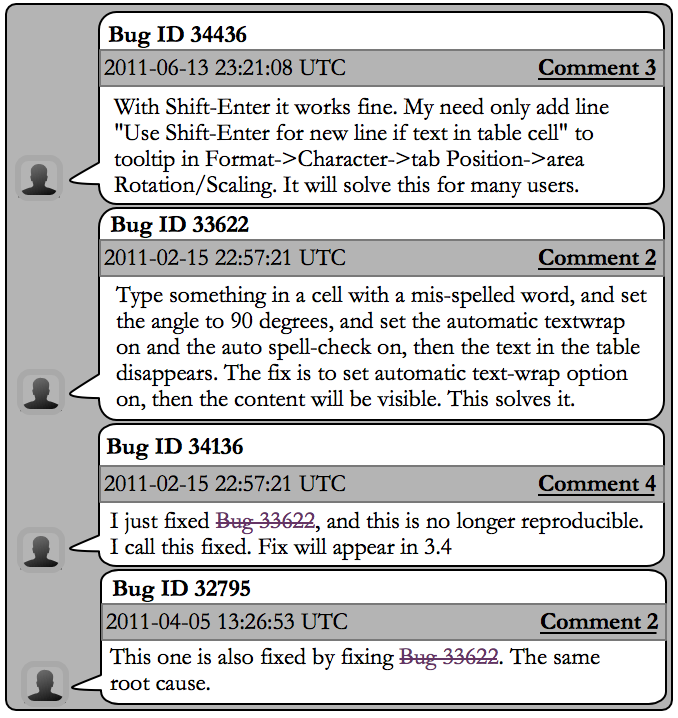}
\caption{\label{second} For the given user query Q$>>$\emph{text cell alignment disappears}, \RR retrieves four relevant bug-fixing comments from various discussion threads of past fixed bugs containing keyword clues in support of bug repair \#34600.}
\label{fig:bug_report_unresolved_b}
\end{center}
\end{figure}
\end{subfigures}

\subsection{Vector Space Model}
\label{sec:VSM}
Vector Space Model (VSM) is a similarity-based ranking model~\cite{salton1975vector}. This model takes each comment and a user query and represents it as an individual document vector $\vec{d}$ of term weights. These weights are determined using term frequency-inverse document frequency TF$\times$IDF~\cite{SaltonB88}. This gives a vector $\vec{d} =$ $(w_{1,d},w_{2,d},...,w_{n,d})$, where $w_{i,d}$ is the term weight for the $i$th term in document $d$. Once the weighted vectors are defined, the cosine similarity metric computes a similarity score. The cosine similarity between two documents $d_1$ (i.e., comments) and $d_2$ (i.e., user query) is computed as follows:
\begin{equation}
sim(d_1,d_2)=\frac{(\sum_{i=1}^n w_{i,d_1} w_{i,d_2})}{\left(\sqrt{\sum_{i=1}^n w_{i,d_1}^2} \cdot \sqrt{\sum_{i=1}^n w_{i,d_2}^2} \right) } 
\end{equation}

\noindent The expression in (Eq.1) generates a similarity score between each comment and a user query input. That means, the higher the ranking score of each comment is, the more likely the user query input is associated with the given comment. 

\subsection{TextRank Model}
\label{sec:textrank_model}
\emph{TextRank} (TR) is a graph-based ranking model~\cite{mihalcea2004textrank}. This model aims to identify relations between various entities in a text by incorporating the concept of \textit{text summarization}~\cite{mitkov, mckeown1995generating,hovy1998automated, salton1999automatic,mani1999advances,radev2002introduction}. The task of text summarization is described in context of finding the relevant sentences in a text and extracting salient terms. The input to the model is a connected graph $G = (V, E)$ with the set of vertices $V$ and set of edges $E$, where $E$ is a subset of V$\times$V. The output of the model is a list of top-N terms with corresponding ranked scores extracted from a corpus. Vertices denote each candidate term. Edges between terms are added between two candidate terms based on their \emph{co-occurrence relation}. The co-occurrence relation indicates the relationships between terms. In our context, we adapted the TextRank graph with undirected edges for each bug comment. An edge is added between two candidate terms in a comment. Then, the weight of the edge is computed where the two words co-occur relative to the number of comments. The TextRank score $TR(V_i)$ associated with a candidate term at vertex $V_i$ is iteratively updated as follows:
\begin{equation}
\small{ TR(V_i)=1-d+d \sum \limits_{V_n \in E(V_i)} \frac {w_{ni}}{ \sum{V_m \in E(V_n) w_{nm}}} TR(V_n) }
\end{equation}

\noindent where $w_{ni}$ is the strength of the connection between two vertices $V_n$ and $V_i, E(V_i)$, and $d$ is the damping factor commonly set to $0.85$~\cite{mihalcea2004textrank, page1999pagerank}. $TR(V_n)$ reports the importance score assigned to a term. The algorithm initially initializes each term with score of $1$. The rank update continues iterating until convergence. The convergence usually ends after $20-30$ iterations. The terms are then sorted based on ranking score. Finally, a list of terms with top-N ranked scores are extracted.

\subsection{Sentiment Analysis}
\label{sec:sentiment_analysis}
\emph{Sentiment Analysis} (SA) which is sometimes referred to as \emph{Opinion Mining}, is a subarea of NLP~\cite{das2001yahoo, morinaga2002mining, pang2002thumbs, tong2001operational,liu2012sentiment}. SA studies people's opinions, sentiments, and attitudes toward entities in a text. SA aims to classify words into two categories: `good' and `bad', and then measure an overall good/bad score in a text. SA can be described as a \emph{classification process}. The goal of SA is to achieve the following tasks: i) find opinions, ii) identify the sentiments they express, and iii) classify their sentiment. Conceptually, this process is designed in four steps. The \textbf{first step} in SA is the \textit{Sentiment Identification}. The input is a text from any domain. The goal is to determine whether the content of the text is \textit{subjective} or \textit{objective}. The \textbf{second step} in SA process is \textit{Feature-Based Classification}. The feature-based classification task treats the text either as bag-of-words (BOWs), or as a string which retains the sequence of words in the document. The \textbf{third step} in SA process is the task of \textit{Sentiment Classification}, which includes machine learning classification approach or lexicon-based approach. Either of these techniques can be applied to one of the three classification levels: document-level, sentence-level, or aspect-level. In our approach, we employ \emph{lexicon-based classification} applied to document-level (i.e., comments). We decided on following this approach because it involves calculating orientation for a bug-fixing comment from the semantic orientation of words or phrases in the comment. Hence, lexicon-based classification relies on a sentiment lexicon, a collection of known and precompiled sentiment terms. It depends on finding the opinion lexicon by using \emph{corpus-based} approach. The corpus-based approach begins with a seed list of opinion words, and then finds other opinion words in a large corpus to help in finding opinion words with context specific orientation. The \textbf{fourth step} in SA process is the \textit{Sentiment Polarity}. The sentiment polarity of a word can be identified by computing the occurrence frequency of the word in a large annotated corpus of texts~\cite{read2009weakly}. If the word occurs more frequently among positive texts, then its polarity is positive. If it occurs more frequently among negative texts, then its polarity is negative. If it has equal frequencies, then it is a neutral word. Our focus is to find the sentiments polarity~\cite{HatzivassiloglouM97} of terms in bug comments as described in the forth step. To achieve this goal, we used NLP python module \textit{TextBlob Sentiment Classifier}~\cite{loria_2016} for processing text. TextBlob takes the text and returns a blob with sentences splitted. Then, built-in sentiment analyzer performs the following functionalities such as extracting terms using Naive Bayes~\cite{mccallum1998comparison} and calculating sentiment for each term using a \emph{lexicon-based score}. The lexicon-based score is obtained by using averaging technique \cite{dave2003mining, turney2003measuring}, in which case each term returns sentiment polarity score. The polarity score is determined within ranges of $[-1.0, 1.0]$, where $1.0$ is highly positive and -$1.0$ is highly negative. An example to a positive comment is:``Try using String class [...] it works fine [...]''.
\begin{figure*}[!t]
\begin{center}
\includegraphics[width=14.8cm]{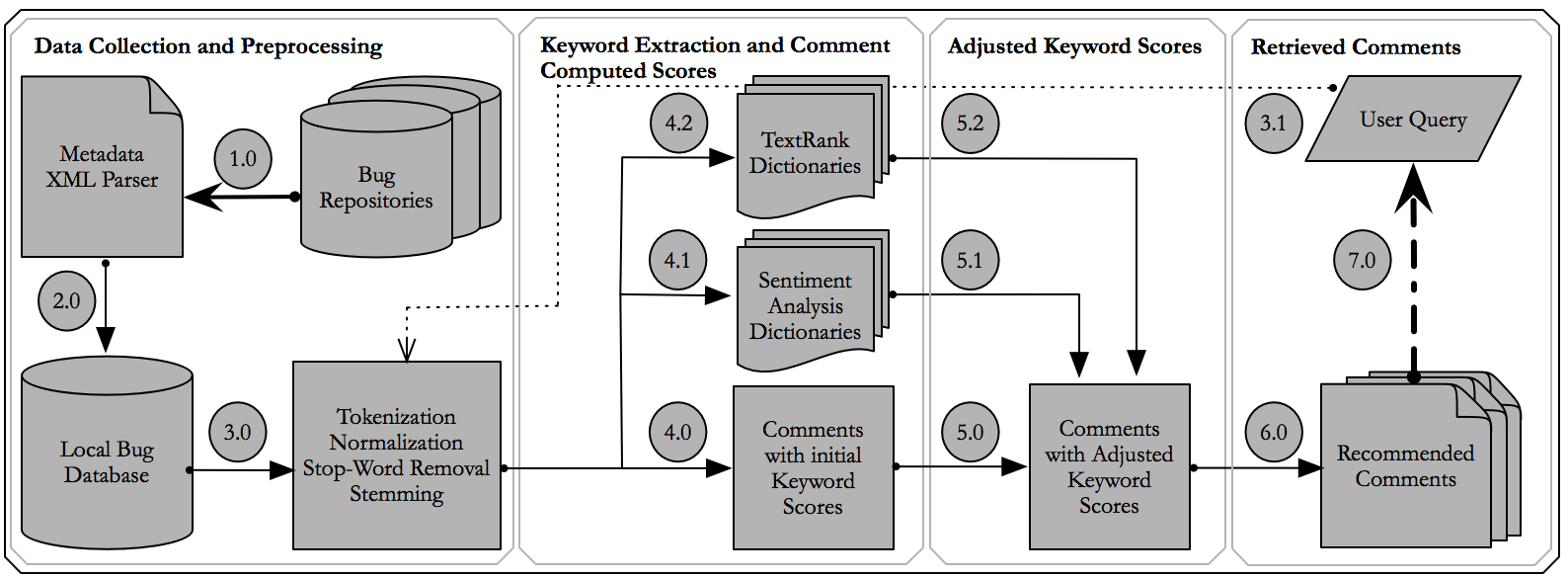}
\vspace{-0.20cm}
\caption{Overview of \RR}
\label{fig:comment_ranking_model}
\end{center}
\end{figure*}

\section{Approach}
\label{sec:approach}
Our ranked-based approach recommends (\RR) a list of comments by leveraging a combined weighted function where keyword weights of each comment are influenced by scores obtained by Vector Space Model (VSM), Sentiment Analysis (SA), and the TextRank Model (TR). 

\subsection{Overview of Architecture}
\label{sec:tool_architecture}
The underlying  architecture of \RR is depicted in Figure~\ref{fig:comment_ranking_model}. The key components are the local database containing \emph{Bug Repositories} (i.e., Bugzilla), the \emph{Comments with Keyword Scores}, the \emph{Sentiment Analysis} that computes sentiment keyword scores based on lookup \emph{bonus/penalty opinion lexicon}, and \emph{TextRank} that ranks keywords based on their co-occurrence and lexical order of importance (i.e., keyword that is most similar to all the other keywords found in comments). \RR is structured into four blocks each representing a set of components. The blocks include: i) Data Collection and Preprocessing, ii) Keyword Extraction and Computed Scores, iii) Adjusted Keyword Scores, and iv) Retrieved Comments. The details are described as follows. \\\vspace{-8pt}

\noindent $\bullet$ \textbf{Data Collection and Preprocessing}: refers to the block represented by the set of components (labels \circled{1.0}, \circled{2.0} and \circled{3.0}, Figure~\ref{fig:comment_ranking_model}). The input to the block are \emph{Bug Repositories} extracted from their XML schema. First, component \emph{Metadata XML Parser} \circled{1.0} parses the metadata. The extracted metadata are: i) Pre-defined attributes of the bug and ii) Free-form text that includes bug comments. Second component, \emph{Local Bug Database} \circled{2.0} saves the metadata that are needed to establish relationships between the bug reports and discussion comments. Third component \circled{3.0}, pre-processes comments into four consecutive steps: i) \emph{Tokenization}, ii) \emph{Normalization}, iii) \emph{Stop-Word Removal}, and iv) \emph{Stemming}. \emph{Tokenization} tokenizes text into terms. \emph{Normalization} aims to eliminate punctuation, non alphanumeric symbols, converting terms to lowercase. \emph{Stop-Words Removal} remove words that carry little meaning. We use a standard NLTK library to perform this task. \emph{Stemming} reduces terms to its root form. For example, terms `crashing' and `crashed' reduce to their base term `crash'. We use the well-known Porter algorithm~\cite{porter1980algorithm} to perform this step. Same preprocessing steps also apply to component \circled{3.1} that represents terms formulated as user query inputs. \\\vspace{-8pt}

\noindent $\bullet$ \textbf{Keyword Extraction and Comment Computed Scores}: refers to the block represented by the set of components (labels \circled{4.0}, \circled{4.1} and \circled{4.2}, Figure~\ref{fig:comment_ranking_model}). Initially, we perform the task of \emph{keyword extraction}~\cite{frank1999domain, turney2000learning}. Keyword extraction selects the keywords that best describe a comment. In our approach, \emph{`Comments with Initial Keyword Scores'} (label \circled{4.0}, Figure~\ref{fig:comment_ranking_model}) is the component where we extract keywords and compute initial keyword weights for each comment and a user query using the TF$\times$IDF schema~\cite{SaltonB88}. In our approach, TF$\times$IDF keywords are words in comments, documents are comments in the bug reports, and the set of documents is the set of all comments in all bug reports. Then we apply the VSM (see Section~\ref{sec:VSM}, Eq.1) to compute the similarity score between each comment and a user query (label \circled{3.1}, Figure~\ref{fig:comment_ranking_model}). Next, we construct the keyword dictionary for \emph{Sentiment Analysis Dictionaries} component (label \circled{4.1}, Figure~\ref{fig:comment_ranking_model}). We incorporate the sentiment analysis (SA) process to keyword extraction as described in Section~\ref{sec:sentiment_analysis}. The process to construct the SA dictionary was based in three steps: i) filter comments across all bugs that are in: \emph{RESOLVED}, \emph{VERIFIED}, for \emph{CLOSED} status, and ii) calculate a `polarity score' for each keyword in each comment, then iii) classify the keywords into two groups: `bonus` and `penalty`. A polarity score is set in the range $[-1.0, +1.0]$~\cite{blair2008building} from which the keyword was extracted $1.0$ for positive, and $-1.0$ for negative. We extract keywords with polarity scores attached to them. Note, we compute the `polarity score' because our goal is to detect positive language in comments. Positive comments are more likely to include clues about the bug than negative ones. For demonstration purposes we compute these comments as follows: POS\_COMMENT: `This bug is fixed' will receive a score of (0+0.05+0+0.8)=0.85, NEG\_COMMENT: `This bug is unresolved' will receive a score of (0+0.05+0-0.8)=-0.75. Next, we construct the keyword dictionary for \emph{TR Dictionaries} component (label \circled{4.2}, Figure~\ref{fig:comment_ranking_model}) based on co-occurrence links between keywords (see Section~\ref{sec:textrank_model}, Eq.2). The process to construct the TextRank dictionary is in four steps: i) filter comments across all bugs that are in: \emph{RESOLVED}, \emph{VERIFIED}, or \emph{CLOSED} status. ii) represent each comment as a graph in which each word is a vertex and each edge is the strength between a pair of vertices. When one vertex links to another, it casts a vote for that vertex. The higher the number of votes that are cast to a vertex, the higher the importance of that vertex, iii) iterate through graph until it convergences using (Eq.2), and iv) finally, sort the vertices based on their final scores and use the values attached to each vertex for ranking.  \\\vspace{-8pt}

\begin{figure*}[!h]
\begin{center}
\includegraphics[width=17.8cm]{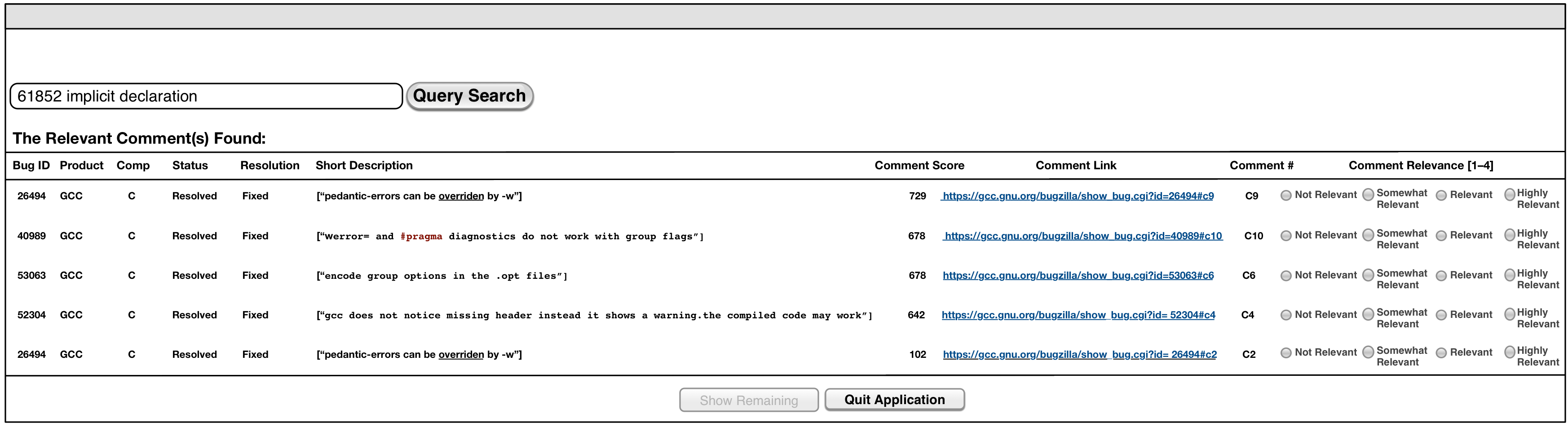}  

\caption{\RR retrieving relevant bug-fixing comments from resolved bugs to address GCC Compiler bug \#61852.}
\label{fig:RetroRank}
\end{center}
\end{figure*}

\noindent $\bullet$ \textbf{Adjusted Keyword Scores}: refers to the block represented by the set of components (labels: \circled{5.0}, \circled{5.1} and \circled{5.2}, Figure~\ref{fig:comment_ranking_model}). The initial scores from VSM (label \circled{4.0}, Figure~\ref{fig:comment_ranking_model}) are then adjusted and boosted by scores of keywords in SA dictionaries (label \circled{5.1}, Figure~\ref{fig:comment_ranking_model}) and by scores of keywords in TR dictionaries (label \circled{5.2}, Figure~\ref{fig:comment_ranking_model}). This combined boosting weighted function $F_w= f(VSM)=SA+TR$ re-ranks each comment's initial VSM keyword scores to more favorable positions in the list influenced by both SA and TR scores. For example, the actual computed final scores for each comment are shown in our tool in Figure~\ref{fig:RetroRank}, column 7 denoted as `Comment Score'.
\\\vspace{-8pt}

\noindent $\bullet$ \textbf{Retrieved Comments}: refers to the block represented by the set of components (labels \circled{3.1}, \circled{6.0} and \circled{7.0}, Figure~\ref{fig:comment_ranking_model}). Given a \emph{User Query}, (label \circled{3.1}, Figure~\ref{fig:comment_ranking_model}), we filter comments with adjusted keyword scores from the previous step in descending order, and iii) finally, we retrieve the top $10$ highest \emph{Recommended Comments} to the user (label \circled{7.0}, Figure~\ref{fig:comment_ranking_model}).

\subsection{Implementation Details}
\label{sec:implementation_details}
We have built a GUI-based tool named \RR as described in Figure~\ref{fig:RetroRank}. \RR takes as an input \emph{a user query} (such as a set of keywords extracted from title and/or long description of bug report). As an output, \RR returns a \emph{ranked list of bug-fixing comments} extracted from cross-cutting discussion threads from past resolved bug. \RR relies on MySQL database where it stores all OSS bug reports and OSS repositories. The UI platform is built upon well-known Python packages: i) \emph{tkinter} that is a standard GUI package to build the framework and ii) \emph{sklearn} used to build functionalities of \RR. In practice, \RR is designed to adapt to any commercial or OSS project that support modern development practices. The inclusion of `pre-merge' code review~\cite{morales2015code, wilson2014best} is an example where relevant comments can be used as part of code review~\cite{halloran2002high, mockus2002two}. Currently, \RR supports four OSS projects (i.e., GCC Compiler, Eclipse IDE, LibreOffice, and Apache Server). A screenshot in Figure~\ref{fig:RetroRank} shows a list of recommended bug-fixing comments that can be used to address bug \#61852. In this example, the \RR recommended several past fixed bugs  (i.e., \#26494, \#40989, \#53063, and \#52304) and their associated bug-fixing comments shown in column 8; 
\section{Evaluation Strategy}
\label{sec:evalstrat}
Our evaluation strategy is to conduct a laboratory \emph{synthetic study} of several configurations (See Section~\ref{sec:synthetic_evaluation}). Then, conduct a larger and more realistic \emph{user study} with human experts comparing the best performing configuration from the synthetic evaluation to the baseline (See Section~\ref{sec:user_study}). The advantage of the synthetic study is that it enables us to test each approach under exactly the same conditions.

\section{Eval-1: Synthetic Study}
\label{sec:synthetic_evaluation}
Synthetic study compares several configurations of our approach to the baseline VSM. In Section~\ref{sec:approach}, we proposed both the Semantic Analysis (SA) and the TextRank (TR) in a combination with the baseline VSM. The objective of the synthetic study is to isolate the effects of both SA and TR: to test whether they improve over the baseline and whether they improve even further when both are combined.

\subsection{VSM Model and its TR and SA Variants}
\label{sec:adapted_tf_idf}

We have used the VSM model implemented by Zhou et \emph{al}.~\cite{zhou2012should} as a baseline to compare our approach. However, the baseline alone does not take into account semantic relations or contextual overlap between keywords or the positive language used among comments. To complement this gap, we introduce the graph-based TextRank model (TR) as a variant of enhancing VSM model. Furthermore, because sentiment patters were noticeable in bug reports~\cite{maalej2015bug,blaz2016sentiment,imtiaz2018sentiment}, we incorporate the Semantic Analysis (SA) to enrich and complement both the VSM model and the TextRank model. Hence, this study comparatively evaluates different variants of approaches. Those are: VSM (baseline), VSM+SA (our approach with TR disabled), VSM+TR (our approach with SA disabled) and VSM+SA+TR (our complete approach, in Section~\ref{sec:approach}).

\subsection{Research Questions}
\label{sec:eval1_rqs}

We ask two research questions towards our objective:
\begin{itemize}
\itemsep0.6em
	\item [RQ$_1$] What is the \emph{ranking performance} of the baseline \emph{VSM}, \emph{VSM+SA}, \emph{VSM+TR}, and \emph{VSM+SA+TR} measured by the metrics we establish in Section~\ref{sec:eval1_metrics}?
	\item [RQ$_2$] Are the effects of \emph{SA} and \emph{TR} orthogonal? \\\vspace{-6pt}
\end{itemize}

The rationale behind RQ$_1$ is to measure the effects of SA and TR individually. Even if VSM+SA+TR outperforms VSM, it is possible that either SA or TR contributes more to the approach. Or, otherwise, preventing the approach from reaching its full potential. Likewise, we ask RQ$_2$ because it is possible that SA and TR provide redundant effects. 

\subsection{Methodology}
\label{sec:eval1_methdodology}
Our methodology to address research questions in the synthetic study is in four steps. First, we manually identify the goldset of bug reports and their corresponding bug-fixing comments; (Note that the goldset could represent bugs that are either in fixed status or duplicates). Second, we generate user queries for each bug report. Third, we use these `user-generated queries' to run each configuration independently. Finally, we compute metrics to measure the performance of each configuration. WE discuss each of the steps as follows: 

\noindent \textbf{Step 1}: To manually identify the goldset, first, we run a script to randomly select the $25$ bugs that were in `FIXED' status from four representative OSS projects as depicted in Table~\ref{tab:open_source_code_projects} and detailed in Section~\ref{sec:subject_projects}. (Note that, we deemed that this number of bugs was sufficient for performing a synthetic study that required extensive manual analysis of goldset data. Likewise, the analysis of the results from multiple combination of approaches (i.e., configurations); This sample was also comparable to other studies~\cite{abal201442,erfani2014works}). Then, for each bug, we identified the bug-fixing comment(s) in the discussion threat that provided the actual fix of a bug---we flagged that comment or a set of comments to be as our goldset. In general, a comment that represents a clue or indicates a bug fix includes the revision link, attached patches, resolution steps or possible links to other duplicate bugs~\cite{noyori2019good,arya2019analysis,zhou2012should}. The revision alone may also include specific file names with their corresponding method signatures and explanations why changes were made. To make the final decision about the goldset, first author and a junior researcher resolved the conflicts of relevance using an established well-known majority--voting mechanism~\cite{zhou2016combining,antoniol2008bug}. Finally, the most highly voted comments were added in the goldset. In Table~\ref{tab:eval1_results}, the goldset is denoted by `Bug ID'. \\\vspace{-8pt}

\noindent \textbf{Step 2:} Then, we recruited $12$ external developers to create/generate the search queries for new bug reports, which do not represent the goldset examples. They are described in Table~\ref{tab:eval1_results}, second column. Developers were asked to read these bug reports and generate as multiple search queries as possible (e.g., extract useful simple keywords from either the bug report title or the bug report long description that best described the root cause of that bug) as though they were searching for a bug-fixing comment. While we could have created these search queries ourselves, we felt that a safer and non-bias approach was to ask developers. Otherwise, we would inadvertently introduce a bias since we had already read the bug report conversations. Each developer read two bugs, with exception of one developer who read three bug reports due to odd number of bugs used for the evaluation. They were instructed to generate as many search queries as possible for that bug report. It should be noted that we only show those `user generated queries' that led to the goldset. The bug reports were not repeated amongst developers. We could have tried multiple developers reading the same bug, but the chances for repeated queries to overlap would have been apparent.  \\\vspace{-8pt}

\noindent \textbf{Step 3}: Next, we used the `user generated queries' from Step 2 as input queries to run each configuration that we were testing (i.e., VSM, VSM+SA, VSM+TR, and VSM+SA+TR). It should be noted that the same `user-generated queries' that were used as inputs to run the baseline VSM were also used to run other configurations under the same conditions. This was necessary to mitigate any bias towards retrieved results. \\\vspace{-8pt}

\noindent \textbf{Step 4}: Finally, once the results were retrieved from each configuration in Step 3, we then marked the ranking position where the goldset (i.e., bug-fixing comments from past resolved bugs) appeared for each configuration retrieved list. The positions are shown in Table~\ref{tab:eval1_results}, columns 3 to 6. Note that there could exist instances where more than one comment from the same bug discussion can be used as goldset as they may relate to the same topic in the conversation. We marked those occurrences with asterisk sign. The criteria of measuring metrics and statistical tests are described in Section~\ref{sec:eval1_metrics}.

\renewcommand{\arraystretch}{0.1}
\begin{table*}[t!]
\caption{Performance between various configurations when `user generated queries' formulated for new bugs are used as inputs for each configuration. \emph{Note that there exist instances of multiple levels of relevance of bug-fixing comments per bug as clues to investigate the new bugs. We denote these instances and their ranking positions in parenthesize and asterisks sign.}}

\vspace{-0.30cm}
\label{tab:eval1_results}
\resizebox{\textwidth}{!}{%
\begin{tabular}{|l|l|c|c|c|c|c|c|}
\hline
OSS Project &
  Q$_i$ \textgreater{} [Bug ID] [keyword$_1\dots$keyword$_n$] &
  VSM &
  VSM+SA &
  VSM+TR &
  VSM+SA+TR &
  \begin{tabular}[c]{@{}c@{}}Goldset \\ (Bug ID)\end{tabular} &
  \begin{tabular}[c]{@{}c@{}}Goldset \\ (Comment ID)\end{tabular} \\ \hline\hline
\multirow{7}{*}{\begin{tabular}[c]{@{}l@{}}GCC GNU \\ Compiler\end{tabular}} &
  Q$_1$ \textgreater \, 60051 unify array domain &
  1 &
  1 &
  1 &
  1 &
  59080 &
  C$_2$ \\ 
 &
  Q$_2$ \textgreater \, 60087 wsign compare warning &
  12 &
  2 &
  2 &
  1 &
  9072 &
  C$_5$ \\ 
 &
  Q$_3$ \textgreater \, 61857 braced-init-list &
  (12, 9)* &
  (5, 3)* &
  (6, 4)* &
  (2, 1)* &
  43875 &
  (C$_3$, C$_2$)* \\ 
 &
  Q$_4$ \textgreater \, 61850 lambda bugs &
  11 &
  4 &
  9 &
  2 &
  56915 &
  C$_2$ \\ 
 &
  Q$_5$ \textgreater \, 61852 implicit declaration &
  (8, 12)* &
  (4, 3)* &
  (6, 4)* &
  (1, 5)* &
  26494 &
  (C$_9$, C$_2$)* \\ 
 &
  Q$_6$ \textgreater \, 78000 wimplicit function declaration macro &
  (10, 12, 15)* &
  (6, 5, 2)* &
  (4, 3, 2)* &
  (3, 1, 2)* &
  71613 &
  (C$_6$, C$_8$, C$_9$)* \\ 
 &
  Q$_7$ \textgreater \, 61151 regression lambda &
  13 &
  6 &
  6 &
  4 &
  47049 &
  C$_{11}$ \\ \hline
\multirow{6}{*}{Eclipse IDE} &
  Q$_8$ \textgreater \, 483049 descriptor &
  5 &
  4 &
  3 &
  3 &
  186451 &
  C$_1$ \\ 
 &
  Q$_9$ \textgreater \, 420612 view menu disabled &
  8 &
  5 &
  3 &
  3 &
  114569 &
  C$_6$ \\ 
 &
  Q$_{10}$ \textgreater \, 493305 text disabled &
  0 &
  0 &
  0 &
  0 &
  -- &
  -- \\ 
 &
  Q$_{11}$ \textgreater \, 494877 selection listener &
  (12, 14, 11, 17)* &
  (6, 5, 4, 2)* &
  (4, 3, 5, 2)* &
  (2, 1, 3, 4)* &
  462760 &
  (C$_{12}$, C$_{13}$, C$_{17}$, C$_{25}$)* \\ 
 &
  Q$_{12}$ \textgreater \, 488771 illegal character parent &
  5 &
  3 &
  4 &
  1 &
  416535 &
  C$_{15}$ \\ 
 &
  Q$_{13}$ \textgreater \, 483849 application mpart &
  8 &
  4 &
  3 &
  1 &
  463962 &
  C$_9$ \\ \hline
\multirow{6}{*}{LibreOffice} &
  Q$_{14}$ \textgreater \, 97465 libreoffice gtk3 &
  3 &
  2 &
  2 &
  1 &
  97419 &
  C$_5$ \\ 
 &
  Q$_{15}$ \textgreater \, 99322 hard recalc &
  (6, 9) &
  (2, 1) &
  (3, 2) &
  (1, 2) &
  89404 &
  (C$_9$, C$_{12)}$* \\ 
 &
  Q$_{16}$ \textgreater \, 33622 rotate text &
  0 &
  0 &
  0 &
  0 &
  -- &
  -- \\ 
 &
  Q$_{17}$ \textgreater \, 97897 recalculate cell rotate &
  6 &
  4 &
  3 &
  1 &
  68034 &
  C$_3$ \\ 
 &
  Q$_{18}$ \textgreater \, 106728 calc cell format &
  (14, 10)* &
  (5, 8)* &
  (6, 3)* &
  (2, 1)* &
  47564 &
  (C$_{12}$, C$_7$)* \\ 
 &
  Q$_{19}$ \textgreater \, 63567 spellcheck cell problem &
  10 &
  3 &
  6 &
  2 &
  37092 &
  C$_2$ \\ \hline
\multirow{6}{*}{Apache Server} &
  Q$_{20}$ \textgreater \, 46214 mod authorization host &
  2 &
  1 &
  2 &
  1 &
  42995 &
  C$_2$ \\ 
 &
  Q$_{21}$ \textgreater \, 33170 mod proxy &
  4 &
  1 &
  2 &
  1 &
  18757 &
  C$_{13}$ \\ 
 &
  Q$_{22}$ \textgreater \, 59019 undefined define issue &
  (5, 8, 7)* &
  (1, 2, 3)* &
  (2, 3, 5)* &
  (1, 2, 3)* &
  35350 &
  (C$_5$, C$_6$, C$_7$)* \\ 
 &
  Q$_{23}$ \textgreater \, 57177 terminated ill signal &
  9 &
  4 &
  3 &
  2 &
  21036 &
  C$_1$ \\ 
 &
  Q$_{24}$ \textgreater \, 33170 remote proxy error &
  (10, 16, 22)* &
  (4, 6, 3)* &
  (5, 9, 6)* &
  (2, 1, 3)* &
  19188 &
  (C$_{12}$, C$_{15}$, C$_{24}$)* \\ 
 &
  Q$_{25}$ \textgreater \, 59136 timeout sent &
  8 &
  5 &
  3 &
  2 &
  58364 &
  C$_1$ \\ \hline\hline
\multirow{3}{*}{\begin{tabular}[c]{@{}l@{}}Metric \end{tabular}} &
  \emph{Average Position of Recommended Comment(s) ($\bm \mu$)} &
  \textbf{9.1} &
  \textbf{3.4} &
  \textbf{3.7} &
  \textbf{1.8} &
  \multirow{3}{*}{} &
  \multirow{3}{*}{} \\ \cline{2-6}
 &
  \multicolumn{1}{r|}{(Top-10) \emph{Mean Average Precision (MAP)}} &
   \textbf{0.192} &
   \textbf{0.428} &
   \textbf{0.358} &
   \textbf{0.741} &
   &
   \\ 
 &
  \multicolumn{1}{r|}{(Top-10) \emph{Mean Reciprocal Rank (MRR)}} &
  \textbf{0.173} &
  \textbf{0.373} &
  \textbf{0.289} &
  \textbf{0.651} &
   &
   \\ \hline
\end{tabular}
}\vspace{-0.25cm}
\end{table*}

\renewcommand{\arraystretch}{0.4}
\begin{table*}[!t]
\caption{Statistical summary of the results from two evaluations. In Eval-1, `n' denotes the $25$ bug reports measured by the \emph{overall ranking performance}. In Eval-2, `n' denotes the number of comments measured by the \emph{overall relevance}. $\mu$ denotes the average position of recommended comment(s) obtained by each configuration. For Eval-1, Student's t-Test results are reported as $p$, $t$, $t_{crit}$ and Cohen's d (effect size). For Eval-2, ANOVA results are reported in terms of $F$, $F_{crit}$, and $p$.}
\label{tab:stat_tests}
\vspace{-0.20cm}
\resizebox{\textwidth}{!}{%
\begin{tabular}{|c|c|c|c|c|c|c|c|c|c|c|c|c|c|c|c|c|}
\hline
\multicolumn{1}{|l|}{Evaluation} &
  H &
  Metric &
  Approach &
  \textit{n} &
  \multicolumn{1}{l|}{\textit{min}} &
  \multicolumn{1}{l|}{\textit{max}} &
  \multicolumn{1}{l|}{$\tilde{x}$} &
  \textit{$\mu$} &
  \textit{$\sigma$} &
  \textit{F} &
  \textit{F\_crit} &
  \textit{p} &
  \textit{t} &
  \textit{t\_crit} &
  \multicolumn{1}{l|}{\textit{Cohen's d}} &
  \textit{Decision} \\ \hline  \hline
\multirow{5}{*}{\begin{tabular}[c]{@{}c@{}}(Eval-1)\\ Synthetic \\ Evaluation\end{tabular}} &
  H$_1$ &
  \begin{tabular}[c]{@{}c@{}}Overall Performance\\ Config-1\end{tabular} &
  \begin{tabular}[c]{@{}c@{}}VSM+SA+TR\\ VSM+SA\end{tabular} &
  \begin{tabular}[c]{@{}c@{}}25\\ 25\end{tabular} &
  \begin{tabular}[c]{@{}c@{}}0\\ 0\end{tabular} &
  \begin{tabular}[c]{@{}c@{}}5\\ 8\end{tabular} &
  \begin{tabular}[c]{@{}c@{}}2\\ 3.5\end{tabular} &
  \begin{tabular}[c]{@{}c@{}}1.8\\ 3.4\end{tabular} &
  \begin{tabular}[c]{@{}c@{}}1.1\\ 1.9\end{tabular} &
  NA &
  NA &
  2.8E-05 &
  -4.8127 &
  2.0301 &
  1.0307 &
  Reject \\ \cline{2-17} 
 &
  H$_2$ &
  \begin{tabular}[c]{@{}c@{}}Overall Performance\\ Config-2\end{tabular} &
  \begin{tabular}[c]{@{}c@{}}VSM+SA+TR\\ VSM+TR\end{tabular} &
  \begin{tabular}[c]{@{}c@{}}25\\ 25\end{tabular} &
  \begin{tabular}[c]{@{}c@{}}0\\ 0\end{tabular} &
  \begin{tabular}[c]{@{}c@{}}5\\ 10\end{tabular} &
  \begin{tabular}[c]{@{}c@{}}2\\ 3\end{tabular} &
  \begin{tabular}[c]{@{}c@{}}1.8\\ 3.7\end{tabular} &
  \begin{tabular}[c]{@{}c@{}}1.1\\ 2.2\end{tabular} &
  NA &
  NA &
  2.9E-06 &
  -5.5575 &
  2.0301 &
  0.1220 &
  Reject \\ \cline{2-17} 
 &
  H$_3$ &
  \begin{tabular}[c]{@{}c@{}}Overall Performance\\ Config-3\end{tabular} &
  \begin{tabular}[c]{@{}c@{}}VSM\\ VSM+TR\end{tabular} &
  \begin{tabular}[c]{@{}c@{}}25\\ 25\end{tabular} &
  \begin{tabular}[c]{@{}c@{}}0\\ 0\end{tabular} &
  \begin{tabular}[c]{@{}c@{}}22\\ 10\end{tabular} &
  \begin{tabular}[c]{@{}c@{}}9\\ 3\end{tabular} &
  \begin{tabular}[c]{@{}c@{}}9.1\\ 3.7\end{tabular} &
  \begin{tabular}[c]{@{}c@{}}5.0\\ 2.2\end{tabular} &
  NA &
  NA &
  1.3E-09 &
  8.1468 &
  2.0638 &
  1.3980 &
  Reject \\ \cline{2-17} 
 &
  H$_4$ &
  \begin{tabular}[c]{@{}c@{}}Overall Performance\\ Config-4\end{tabular} &
  \begin{tabular}[c]{@{}c@{}}VSM\\ VSM+SA\end{tabular} &
  \begin{tabular}[c]{@{}c@{}}25\\ 25\end{tabular} &
  \begin{tabular}[c]{@{}c@{}}0\\ 0\end{tabular} &
  \begin{tabular}[c]{@{}c@{}}22\\ 8\end{tabular} &
  \begin{tabular}[c]{@{}c@{}}9\\ 3.5\end{tabular} &
  \begin{tabular}[c]{@{}c@{}}9.1\\ 3.4\end{tabular} &
  \begin{tabular}[c]{@{}c@{}}5.0\\ 1.9\end{tabular} &
  NA &
  NA &
  2.0E-09 &
  8.0048 &
  2.0638 &
  1.5071 &
  Reject \\ \cline{2-17} 
 &
  H$_5$ &
  \begin{tabular}[c]{@{}c@{}}Overall Performance\\ Config-5\end{tabular} &
  \begin{tabular}[c]{@{}c@{}}VSM\\ VSM+SA+TR\end{tabular} &
  \begin{tabular}[c]{@{}c@{}}25\\ 25\end{tabular} &
  \begin{tabular}[c]{@{}c@{}}0\\ 0\end{tabular} &
  \begin{tabular}[c]{@{}c@{}}22\\ 5\end{tabular} &
  \begin{tabular}[c]{@{}c@{}}9\\ 2\end{tabular} &
  \textbf{\begin{tabular}[c]{@{}c@{}}9.1\\ 1.8\end{tabular}} &
  \begin{tabular}[c]{@{}c@{}}5.0\\ 1.1\end{tabular} &
  NA &
  NA &
  \textbf{3.0E-11} &
  9.5131 &
  2.0638 &
  \textbf{2.0165} &
  Reject \\ \hline \hline
\begin{tabular}[c]{@{}c@{}}(Eval-2)\\ User Study\\ Evaluation\end{tabular} &
  H$_6$ &
  Overall Relevance &
  \begin{tabular}[c]{@{}c@{}}VSM+SA+TR\\ VSM\end{tabular} &
  \begin{tabular}[c]{@{}c@{}}410\\ 597\end{tabular} &
  \begin{tabular}[c]{@{}c@{}}1\\ 1\end{tabular} &
  \begin{tabular}[c]{@{}c@{}}4\\ 4\end{tabular} &
  \begin{tabular}[c]{@{}c@{}}3\\ 1\end{tabular} &
  \begin{tabular}[c]{@{}c@{}}\textbf{2.7}\\ \textbf{1.4}\end{tabular} &
  \begin{tabular}[c]{@{}c@{}}1.07\\ 0.77\end{tabular} &
  488.31 &
  3.85 &
  \textbf{4.5E-75} &
  -20.818 &
  1.9634 &
  \textbf{2.9456} &
  Reject \\ \hline
\end{tabular}%
}
\end{table*}

\subsection{Participants}
\label{sec:eval1_participants}
For step 2 in the previous section, we invited developers from three mid-sized software development companies by sending them an invitation survey. We categorized participants according to their level of experience. We selected $12$ participants. Participants helped to create a dataset for the synthetic study. The participants did not offer subjective evaluations as they did for the user study. Also it should be emphasized that the participants from synthetic study and user study did not overlap. We view this as an advantage because it means a more diverse base of participants, which should reduce potential biases. All participants had more than five years programming experience. The preliminary online survey showed that about $91.7$\% of them were skillful in both C and C++. All of them reported high proficiency in Java. About $58.3$\% of them were familiar with the GCC Compiler. The majority of them, about $91.7$\% were familiar with Apache Server. While, $41.7$\% of them were familiar with LibreOffice, all of them knew Eclipse.

\subsection{Metrics and Statistical Tests}
\label{sec:eval1_metrics}

The metrics to evaluate research questions RQ$_1$ and RQ$_2$ are established based on \emph{ranking} and \emph{orthogonality}.

\begin{itemize}
\itemsep0.4em
	\item [RQ$_1$] \emph{\textbf {Ranking}}: To compute the ranking, the procedure was two fold: i) we counted the position where the goldset (i.e., identified bug-fixing comments) appeared in the retrieved list across all configurations, and ii) we compared their ranking position to evaluate which of the configurations performed better. It should be noted that the higher the position of the goldset comment, the better the performance of that particular configuration.
	
	\item [RQ$_2$] \emph{\textbf{Orthogonality}}: Similar to procedure described in RQ$_1$, we evaluated the ranking differences of the results returned from our approach VSM+SA+TR against ranking differences of the results returned from variants VSM+SA and VSM+TR. This was necessary to investigate the effect of TR and SA impose on the performance between our approach and the baseline VSM.
\end{itemize}

\noindent \emph{Student's t--Test}~\cite{smucker2007comparison} was used as a metric similar to other studies~\cite{GouesDFW12,WangLJLL11,bettenburg2008duplicate} to evaluate the statistical significance of differences in \emph{ranking} and \emph{orthogonality} of each configuration. We assumed on the \emph{null} hypothesis that the mean values of the impact set sizes for all configuration were equal. Additionally, we leveraged standard ranked-based metrics similar to other studies~\cite{zhou2012should,rath2018analyzing,gharibi2017locating} to enable a clear-cut comparisons among various approaches/configurations. Specifically, we measured \emph{mean average precision (MAP)} and \emph{mean reciprocal rank (MRR)}~\cite{manning2010introduction}. MAP considers average rank position of each relevant comment; MRR considers average position of the first relevant comment in the ranked list.
\section{Eval-1: Synthetic Study Results}
\label{sec:eval1results}
\subsection{RQ1: Overall Ranking Performance}
We found that VSM+SA+TR achieved the highest performance in terms of ranking position of the recommended bug-fixing comments, followed by VSM+TR, VSM+SA and the baseline VSM. In Table~\ref{tab:eval1_results}, H$_5$, the average ranking position ($\mu$) of the recommended bug-fixing for Config-5: VSM+SA+TR was ($1.8$) and for the baseline Config-5: VSM it was ($9.1$). In other words, on average, VSM+SA+TR placed the bug-fixing comments around the second or the first position, whereas the baseline VSM placed it in the ninth position. We also noticed that Config-4: VSM+SA and Config-3: VSM+TR placed the bug-fixing comments in a higher position than the baseline VSM for those configurations. In two cases, no approach located the bug-fixing comments. This most likely occurred due to no past comments found relevant to bug reports: \#493305 or \#33622. In only one case did the baseline VSM placed the recommended bug-fixing comment in the first position. VSM+SA+TR placed the recommended bug-fixing comment in the first position for $16$ instances of bug report queries. Results indicate that differences in average ranking position $(\mu)$ of the bug-fixing comments are statistically significant as reported in hypothesis H$_5$ ($3.0$E$-11$), Table~\ref{tab:stat_tests} with p-value of $0.05$ and a confidence level of $0.95$ confirming the statistical significance. The Cohen's d value~\cite{rice2005comparing} indicates a large effect size ($d=2.0165>0.8$). That said, approaches by hypothesise H$_5$ are highly significant in terms of rankings. Similarly, we observe that both MRR and MAP scores are higher for VSM+SA+TR than VSM+TR or VSM+SA, and significantly higher than scores obtained for the baseline VSM. What that means is that our approach (VSM+SA+TR) significantly improves retrieval performance by 74\% in terms of mean average precision (MAP) as denoted in Table~\ref{tab:eval1_results}. 

\subsection{RQ$_2$: Orthogonality}
To answer RQ$_2$, results indicated that both SA and TR provided orthogonal information. We observed that VSM+SA and VSM+TR outperformed the baseline VSM; In some instances, VSM+SA outperformed VSM+TR and vice-versa. If the approaches did not provide orthogonal information, we would expect that both approaches VSM+SA and VSM+TR most likely would \emph{not} improve VSM+SA+TR. For example, we noticed that for queries Q$_3$, Q$_4$, Q$_{14}$, Q$_{15}$, Q$_{19}$, Q$_{20}$, Q$_{22}$, and Q$_{24}$, VSM+SA provided higher improvement than VSM+TR. Consider the example of Q$_{15}$, VSM+SA returns two comments (i.e., (C$_9$, C$_{12}$)*) from a same bug one position above (i.e., (2, 1)*) compared to VSM+TR which returns the same comments one position below for each comment (i.e., (3, 2)*). That said, the recommended bug-fixing comment that were found during the query search  (C$_9$, C$_{12}$)* from bug \#89404 were considered as candidate clues for addressing the LibreOffice bug \#99322 and it was slightly better ranked by VSM+SA than VSM+TR. If we closely examine one of the recommended comment e.g., \textbf{C$\bm{_9}$} for query Q$_{15}$ in the boxed area below, it consists of words that are indicative of TR and SA representations. As noted in the text-box, VSM+SA has picked \emph{more} positive words that are shown in bold: `resolved', `fixed', `fine', `right', `IMO'. In contrast, VSM+TR has picked \emph{only a few} important lexical units that are underlined as: `saving and reopening' and `hard recalc'.  \\\\\vspace{-0.40cm}

\noindent \fbox{\begin{minipage}{24.5em}
\textbf{C$\bm{_9}$}: Doing ONE \underline{hard recalc} both files work \textbf{fine} now... see Build ID c4c7d32d0d49397cad38d62472b0bc8acff48dd6 and linked patch \underline{saving and reopening} after \underline{hard recalc} keep the \textbf{right} values...There was several \textbf{fixed} for 4.4.0, \textbf{fixed} now. \textbf{IMO} this is \textbf{resolved} as worksform."
\end{minipage}} \\\\\vspace{-0.20cm}

\noindent These keyword provides positive affirmation for recommending clues to resolve a bug. Besides the positive keywords, the comment \textbf{C$\bm{_9}$} provides other information such as build link ID, patch link that are useful for bug localization.
\section{Threats to Validity (Eval-1)}
\label{sec:eval1_threats}

In the synthetic study, we identified three main threats to validity. The first threat is the use of $25$ bug reports provided in the synthetic study as denoted in Table~\ref{tab:eval1_results}. We could have completely omitted this step and only use the best performing configurations for the user study. However, our goal was to introduce a mixed-method to better assess the quality of different configurations used in the study. We followed a \emph{Goal-Metric-Question Paradigm}~\cite{basili1992software} that is rarely practiced to answer research questions within the scope of bug localization. The second threat is the tasks themselves. We selected real bug reports from four well-known open-source repositories. The representation of bugs carry various complexities ranging from [P$_1$--P$_4$] priority bug reports. Often times these prioritized bug reports also require expert-knowledge. The query entries formulated by the participants could return content outside the scope of actual bug description. We attempted to mitigate this gap by allowing participants to formulate as many queries as possible when they were unsure about the requirements of the bug reports. The third threat is related to complexity of bug reports. Even though the majority of participants had more than five years of industry experience, there were certain bug reports that required domain knowledge. For example, a few participants struggled with specific GCC Compiler terminologies. We attempted to mitigate this threat by relying on our background questionnaire. We specifically asked participants about their expertise prior to completing any of their tasks.
\section{Eval-2: User Study}
\label{sec:user_study}

In the user study, we compare the best-performing approach VSM+SA+TR to the baseline VSM~\cite{zhou2012should}. Our objective is to measure comment(s) relevance returned by each approach.

\subsection{Research Question}
\label{sec:rq3}

We ask the following research question to address relevance:

\begin{itemize} \setlength\itemindent{8pt}
\item[RQ$_3$:] To what extent do comments retrieved by the baseline \emph{VSM} and \RR that combines \emph{VSM+SA+TR} differ in \emph{relevance} as perceived by the experts? \\\vspace{-6pt}
\end{itemize}

\noindent Research question RQ$_3$ is intended to fill a gap that is left by the synthetic study. In the synthetic study, we found that when \RR combines VSM+SA+TR bug-fixing comment are substantially ranked in a better position than rankings obtained with VSM only. However, in the top-10 results from a search tool, all results may have some degree of relevance to the search query. So even though the baseline places the recommended answer at position $10$, the first $9$ results may provide useful information for developer to investigate the bug. Intent for RQ$_3$ is to measure the perceived relevance of the top-$10$ results of each tool--otherwise we could not be sure that one tool is outperforming the other.

\begin{table}[t!]
\renewcommand{\arraystretch}{1.0}
\centering
\caption{Cross-validation design of the \emph{user study}.}
\label{tab:xval}
\resizebox{\columnwidth}{!}{%
\begin{tabular}{|c|c|c|c|}
\hline
Experiment & Group & Search Tool & Task Set \\ \hline \hline
\multirow{2}{*}{1} &A& VSM+SA+TR (\RR) &Bug01--Bug20 \\ 
&B& VSM (baseline) & Bug01--Bug20 \\ \hline
\multirow{2}{*}{2} & A & VSM (baseline) & Bug21--Bug40 \\  
& B & VSM+SA+TR (\RR) & Bug21--Bug40 \\ \hline

\end{tabular}
} 
\end{table}

\begin{table*}[!t]
\renewcommand{\arraystretch}{1.1}
\centering
\caption{Characteristics of the open-source projects}

\label{tab:open_source_code_projects}
\resizebox{\textwidth}{!}{\begin{tabular}{|l| l | c || c | c | c | c | c |} \hline
OSS Project   & \multicolumn{1}{c|}{Project Description}                            & \begin{tabular}[c]{@{}c@{}}Interval \end{tabular} & Product Name & Component Type & \#Bug Reports   & \# Resolved Bugs & \# Unresolved Bugs \\ \hline \hline
GCC GNU Compiler  & Compiler system produced by the GNU project & Aug 1999 -- Aug 2018& GCC & C and C++ & 75901 & 66608 & 9293               \\
Eclipse IDE & Development platform for Java language  & Mar 2002 -- May 2017 & Platform  & IDE & 8028  & 4321 & 3707               \\
LibreOffice & Office used for document creation & Oct 2010 -- May 2018 & LibreOffice & CALC & 47336 & 1757 & 27540              \\
\textit{Apache Server} & Web server platform on the HTTPd server& Jan 2002 -- Apr 2019 & HTTPD-2 & CORE & 2511  & 19796  & 754                \\ \hline \hline
\multicolumn{5}{|r|}{\textit{Total \#Bug Reports:}}                                      & \textbf{133776} & \textbf{92482}   & \textbf{41294}    \\ \hline
\end{tabular}
}
\end{table*}

\subsection{Methodology}
\label{sec:eval2_methodology}
Our methodology for the user study was to ask participants to read bug reports, then use one of the approaches (VSM or VSM+SA+TR) to search for similar related bug-fixing comment(s) from past resolved bugs. To avoid biases, we used a cross-validation design to evaluate $40$ real bug reports. Table~\ref{tab:xval} shows this design of two separate experiments. Prior to each experiment, we recruited $20$ participants (\emph{not} overlapping with the participants recruited to create a goldset in the synthetic study), and randomly divided these participants into two groups of $10$ (denoted A and B). Participants were prevented from knowing whether they were evaluating our approach or the baseline approach to avoid introducing bias. Both tools were identical in both scenarios even in terms of UI look-and-feel features. During the study our approach was denoted as `Tool A' and the baseline approach as `Tool B'. Once the tool naming situation was consolidated we asked in experiment 1, participants in group A to use VSM+SA+TR (a.k.a `Tool A') to search for bug-fixing comments for bugs 1-20. Participants in Group B used the baseline VSM (a.k.a `Tool B') for the same bug reports. Then, in experiment 2, group A used VSM and group B used VSM+SA+TR, but on a different set of $20$ bug reports. The end effect was that different participants used different tools for different bug reports. During each experiment, a participant would read a bug report, formulate a query, and enter that query into the tool. Each participant was given one hour interval for each experiment to complete the bugs in each group. The participants then rated each of the top--$10$ results from each tool on a scale from $[1-4]$ as we will describe in Section~\ref{sec:eval2_metrics}. Figure~\ref{fig:RetroRank} depicts the rating scale each participant would select depending on which comment would be more relevant to addressing the new bug report. The cross-validation design follows the norms established in software engineering. We asked participants to evaluate the top--$10$ results due to the accepted practice that researchers follow~\cite{granka2004eye,ye2016mapping, rath2018analyzing}.

\subsection{Participants}
\label{sec:participants}
We recruited $20$ participants to conduct the user study. Half of the participants in the study had industry experience. The rest of the participants were recruited from academia. Based on a preliminary online survey, $81$\% of the participants had more than $3$ years of programming experience. From all participants, $90.5$\% of them were skillful in C, C++, and Java. About $40$\% of their current job was in software development. In addition, $57.1$\% of the them asserted their experience with issue-tracking systems. Participants distributed their experience among four projects as follows: i) $52.4$\% of them were familiar with GCC Compiler, ii) $85.7$\% were familiar with Eclipse IDE, iii) $33.3$\% were familiar with Apache Server, and iv) Lastly, $42.9$\% of them were familiar with LibreOffice.

\subsection{User Study Tasks}
\label{sec:tasks}
We designed the experiments by providing $40$ bug reports as the main tasks for the user study. To avoid introducing any bias in the study, the bug reports were real-world examples randomly selected from four distinct open source projects as reported in Table~\ref{tab:open_source_code_projects} and Section~\ref{sec:subject_projects}. All $40$ bugs were identical to the ones found via online issue-tracking systems. These bugs varied in terms of priority [$P_1$-$P_4$]. In bug report prioritization, $P_1$ and $P_2$ are categorized as high priority bugs whereas $P_3$ and $P_4$ are categorized as low priority bugs that do not require immediate fix. 

\subsection{Subject Open-Source Projects}
\label{sec:subject_projects}
The bug repositories of open-source projects are considered the main source to evaluate \RR and the baseline. Table~\ref{tab:open_source_code_projects} reports four different open-source software systems. Those are: \emph{GCC GNU Compiler}, \emph{Eclipse IDE}, \emph{Apache Server}, and \emph{LibreOffice}. All four projects include a full repository of bug reports filtered by \emph{product} and \emph{component} type. Our decision to select these four projects is based on four premises. First, all projects fall within three categories of small, medium, and large projects. Hence, they are generalizable across domains. Second, all projects use bug tracking systems and version control to easily trace code changes.  Third, all projects consist of bug reports containing lengthy discussions with hundreds of comments. Finally, they are extensively used in software engineering research community.

\begin{table*}[!t]
\renewcommand{\arraystretch}{1.1}
\centering
\caption{Retrieval and relevance scores of comments returned by both our approach VSM+SA+TR and the baseline VSM. These results are reported from the user study completed by two participants chosen from two groups to address Bug ID:61852.}
\label{tab:relevance_scores_results}
\vspace{-0.20cm}
\resizebox{\textwidth}{!}{%
\begin{tabular}{|l|l|l|l|l|l|c|}
\hline
Approach & Project & Short Description & User Query Input & Query Output Results & Bug \#, Comment & \multicolumn{1}{l|}{Relevance Score {[}1-4{]}} \\ \hline\hline
\multirow{4}{*}{VSM+SA+TR} & \multirow{4}{*}{GCC} & \multirow{4}{*}{\begin{tabular}[c]{@{}l@{}}Incorrect column number for \\ -Wimplicit-function-declaration\end{tabular}} & \multirow{4}{*}{\begin{tabular}[c]{@{}l@{}}Wimplicit function \\ declaration\end{tabular}} & pedantic info can be overridden by -w & 26494, $C_9$ & 4 \\ \cline{5-7} 
 &  &  &  & \begin{tabular}[c]{@{}l@{}} werror=and \#pragma diagnostics \\ should work with group flags\end{tabular} & 40989, $C_{10}$ & 3 \\ \cline{5-7} 
 &  &  &  & \begin{tabular}[c]{@{}l@{}}yes definitely encode group options \\ in the .opt files\end{tabular} & 53063, $C_6$ & 3 \\ \cline{5-7} 
 &  &  &  & \begin{tabular}[c]{@{}l@{}}gcc should use header instead it shows \\ a warning.the compiled code is fixed now\end{tabular} & 532304, $C_4$ & 4 \\ \hline \hline
\multirow{4}{*}{VSM} & \multirow{4}{*}{GCC} & \multirow{4}{*}{\begin{tabular}[c]{@{}l@{}}Incorrect column number for \\ -Wimplicit-function-declaration\end{tabular}} & \multirow{4}{*}{\begin{tabular}[c]{@{}l@{}}Wimplicit function \\ declaration\end{tabular}} & \begin{tabular}[c]{@{}l@{}}internal compiler error in rest\_of\_compilation\\ at toplev.c:3491\end{tabular} & 9738, $C_9$ & 1 \\ \cline{5-7} 
 &  &  &  & the compiler crashes & 5830, $C_{10}$ & 1 \\ \cline{5-7} 
 &  &  &  & \begin{tabular}[c]{@{}l@{}}3.3 regression mingw internal compiler error \\ in rest\_of\_compilation, at toplev.c:3491\end{tabular} & 9738, $C_9$ & 2 \\ \cline{5-7} 
 &  &  &  & undefined reference to non-virtual thunk to... & 4122, $C_6$ & 1 \\ \hline
\end{tabular}%
}
\end{table*}

\subsection{Metrics and Statistical Tests}
\label{sec:eval2_metrics}
\noindent The metric to evaluate the research question RQ$_3$ is as follows:
\begin{itemize}
\setlength\itemindent{4pt}
\item [RQ$_3$] \emph{\textbf{Relevance}}:
Comments were evaluated by the participants on a four-point Likert scale from [1-4] where four is `highly relevant', three is `relevant', two is `somewhat relevant' and one is `not relevant'. To compute the metric for $RQ_3$, we used the comment relevance score metric. \\\vspace{-8pt}
\end{itemize}

Our criteria for selecting comment relevance is as follows: \\\vspace{-8pt}

\begin{enumerate}
	\itemsep0.3em
	\item \emph{Not Relevant} (1)---the information the comment contains does not include any relevant information to query input.
	
	\item \emph{Somewhat Relevant} (2)---the information the comment contains, mostly lacks details to references to other artifacts or links, the explanations may be useful for partial understanding or limited bug investigation clues.
	
	\item \emph{Relevant} (3)---the information the comment contains, reveals details such as: artifact links that trace the bug, and to some extent, commit patches with partial explanations.
	
	\item \emph{Highly Relevant} (4)---the information the comment contains reveals extremely useful details such as: code snippets and patches with diffs, links to related/duplicate reports, resolution steps and change explanations.
\end{enumerate}
\section{Eval-2: User Study Results}
\label{sec:eval2results}
This section describes results from our user study, as well as our conclusions and answer to RQ$_3$.\\ \vspace{-8pt}

\noindent \emph{\textbf{RQ${\bm{_3}}$: Overall Relevance}}. Results show that \RR when using VSM+SA+TR achieved higher relevance than when the baseline VSM was used alone~\cite{zhou2012should}.The evidence is reported in Table~\ref{tab:stat_tests}. \RR (VSM+SA+TR) based on statistical significance reported by hypothesis H$_6$. Note that the obtained statistical test results are based on metrics defined in Section~\ref{sec:eval2_metrics}. A concrete example is also described in Table~\ref{tab:relevance_scores_results}. Due to space limitations we only show results obtained by two participants. The key data in Table~\ref{tab:relevance_scores_results} is the `Relevance Score' column which shows the ranking scores from $[1-4]$ as defined in Section~\ref{sec:eval2_metrics}. The relevance scores that participant P$_{9}$ in (group A) selected for \RR (VSM+SA+TR) were significantly higher than the relevance scores that participant P$_9$ in (group B) selected for VSM~\cite{zhou2012should}. We also observe a universal pattern across all bugs with the relevance scores from both approaches. For example, for the baseline VSM, the relevance scores tend to decrease within the top-10 query but for \RR (VSM+SA+TR) tend to increase within the top-10 query results. Another evidence found in this data is that the recommended comment that is placed in the first position is marked as `highly relevant'. However, we do not perceive a similar outcome from the results returned by the baseline VSM. Additionally, the query results from the baseline do not show a strong relationship relevant to bug's content in (group B). The query results from \RR (VSM+SA+TR) show a stronger relationship with the content from the query input and bug description. To illustrate, in all cases where relevance score is either $3$ or $4$, the results consist of vocabulary that refer to compiler signaling warnings. Hence, we pose that the differences in relevance of comments are statistically significant between \RR (VSM+SA+TR) and the baseline VSM. Such evidence is provided in Figure~\ref{tab:stat_tests}, hypothesise H$_6$, ($4.5E-75$) with p-value of $0.05$ and a confidence level of $0.95$. The average relevance scores for \RR (VSM+SA+TR) ($\mu=2.7$) are higher compared to the average relevance scores for the baseline VSM ($\mu=1.4$). 
\section{Threats to Validity (Eval-2)}
\label{sec:threats_to_validity_2}
\noindent There are several threats that impact the validity of our work. We focus on internal and external threats. \\\vspace{-6pt}

\noindent\textit{\textbf{Internal Threats:}} The internal threat of a study is the extent to which a treatment effects change in the dependent variable. In this experiment it could be due to the content of the comment. If a comment lacks informativeness~\cite{Bettenburg2007,zhang2017bug}, the outcome of \RR could be affected. Other factors may relate to individual differences among subject groups. A few participants more than the others may be influenced by factors of stress, tiredness or programming experience. We attempt to minimize these threats by alternating the order of tasks at each participant's turn using a cross-validation design. We further instruct subjects not to discuss the experiments with each other. \\\vspace{-6pt}

\noindent \textit{\textbf{External Threats:}}
The external threat of a study is the extent to which we can generalize the results of our user study. Our evaluation focused solely on OSS bug reports. A threat that might limit the generalizability of our results is the use of only $40$ real-world bug reports. We attempt to minimize this representative sample by selecting bug reports from four well-known open source domains. To mitigate further threats, we also engaged subjects from academia and industry to help generalizing our findings to both contexts.
\section{Related Work}
\label{related_work} 

Current studies have developed various methods that focus on automatically ranking source files for given bug reports. Kim et \emph{al}.~\cite{KimTKZ13} applied a two-phase machine learning approach to predict files that could fix a bug report. DebugAdvisor~\cite{AshokJLRSV09} used searches over structured and unstructured text to recommend a set of similar past tasks and the files that were changed. Saha et \emph{al}.~\cite{SahaLKP13} presented BLUiR and used the structure of bug reports and source code files to build a retrieval structure for reporting the bugs. Nguyen et \emph{al}.~\cite{NguyenNANN11} applied LDA model to predict buggy files for given bug reports. Similarly, Lukins et \emph{al}.~\cite{LukinsKE10} applied LDA to predict and filter the top $10$ defective files across $22$\% of overall bug reports. Ye et \emph{al}.~\cite{YeBL14} used developer domain knowledge (e.g., API documents) to identify source files associated to a bug. The same authors proposed a ranking model to recommend top $10$ source files~\cite{ye2016mapping}. Zhou et al.~\cite{zhou2012should} used VSM model to recommend textual similarities among similar past bugs and their associated files to address a new bug. Our approach instead recommends similar past bugs by extracting relevant bug-fixing comments from lengthy discussion threads of past bugs to address a new bug. A line of research by Runeson et \emph{al}.~\cite{RunesonAN07}, Wang et \emph{al}.~\cite{WangZXAS08}, Bettengurg et \emph{al}.~\cite{bettenburg2008makes}, Sun et \emph{al}.~\cite{SunLKJ11}, Hindle and Onuczko~\cite{hindle2019preventing} focused on detecting duplicate reports by querying bug reports.
\section{Conclusion and Future Work}
\label{sec:conclusion}
In this paper, we introduced a novel approach for recommending related bug-fixing comments from cross-cutting discussion threads of past solved bugs in support of bug repair. It is a common practice that when the developers investigate a new bug they often need to know how similar bugs were solved in the past. Our approach was based on both a combination of Sentiment Analysis and the TextRank semantic-based techniques. By incorporating both combinations into a text similarity strategy (VSM), we identified bug-fixing comments that were more positive and semantically connected. Additionally, two evaluations demonstrated that both types of information improved search, relevance. It also provided insightful summary of bug repair instructions. For future work, our goal is to perform a systematic qualitative study where the rationale behind the achieved results will be shown with real scenarios of bug-fixing examples explained in greater detail. 

\bibliographystyle{abbrv}
\balance \footnotesize \bibliography{bibliography.bib}

\end{document}